\documentstyle[preprint,aps,floats]{revtex}
\begin{document}
\title{SUPERPOSITION OF A STATIC PERFECT FLUID AND A RADIAL 
ELECTRIC FIELD\thanks{gr-qc/9605040}}
\author{Mauricio Cataldo\thanks{e--mail: mcataldo@zeus.dci.ubiobio.cl}
\\{\small Departamento de F\'\i sica, 
 Facultad de Ciencias, Universidad del B\'\i o-B\'\i o, Concepci\'on, Chile}\\
Patricio Salgado\thanks{e--mail: psalgado@halcon.dpi.udec.cl}
\\{\small Departamento de F\'\i sica, Facultad de Ciencias,
Universidad de Concepci\'on, Concepci\'on, Chile}}
\maketitle
\begin{abstract}
We obtain a two-parameter set of solutions, which represents a spherically
symmetric space-time with a superposition of a neutral fluid and an electric
field. The electromagnetic four-potential of this Einstein-Maxwell space-time
is taken in the form $A=\frac{q}{n}r^{n}dt$, when $n \neq 0$ and 
$A=q\,ln\, r\,dt$, when n=0
(where $q$ and $n$ are arbitrary constants). Some particular solutions
obtainable from the general solution are presented.
\end{abstract}

\section{Introduction} 
In this paper we formulate an approach that gives us a procedure to write
down in an easy way the Einstein-Maxwell equations and solve them for a metric
with certain symmetries. This procedure is formulated on the basis of a 
generalization of the approach to generation of Einstein-Maxwell fields
proposed by J. Horsk\'y and N.V. Mitskievitch~\cite{Horsky1} and later 
developed in~\cite{CataldoA,Cataldo1,Horsky2,Tsalakou,Cataldo2,Cataldo3},
where Einstein-Maxwell exact solutions are generated from a seed isometric 
gravitational field. In particular, in~\cite{Horsky1} is formulated a 
conjecture about the connection between isometries of {\bf vacuum} space-time 
and the existence of corresponding space-times with electromagnetic fields for 
which the electromagnetic four-potential is proportional to a Killing vector 
of the vacuum seed metric. Some electrovacuum solutions are obtained 
in~\cite{CataldoA,Cataldo3}. The new method of generation of electrovacuum 
solutions was later generalized to the case of non-vacuum seed solutions. Some 
superpositions of electrically neutral stiff matter with an electric field are 
obtained in~\cite{Cataldo1,Horsky2}. In this case the four-potential must be
proportional to a Killing vector which is orthogonal to the four-velocity
of the neutral fluid in the seed gravitational field. Self-consistent fields 
with a charged perfect fluid are generated in~\cite{Tsalakou,Cataldo2}. We 
will show that one can leave out the seed solution 
and consider only the chosen form of the self-consistent line element and 
four-potential, from which we obtain all the required parameters for the 
electromagnetic field.

On the other hand, we apply our approach to the construction of a 
superposition of a radial electric field and a static neutral perfect
fluid, based on the idea of the re-interpretation of the exterior 
electrovacuum solution of Kottler-Reissner-Nordstr\"{o}m~\cite{Kramer,Cataldo2}
as a superposition of a coulombian type electric field and a static perfect 
fluid with
\begin{math}
\mu = -p = \Lambda/\chi, 
\end{math}
where $\Lambda$ is the cosmological constant, $\mu$ and $p$ are the energy density 
and pressure of the fluid, respectively. In Sec. II we discuss the main ideas 
of the construction of the above mentioned superposition for the space-time 
with the most general static spherically symmetric line element. In Sec. III 
we obtain our two-parameter set of metrics for the electromagnetic 
four-potential determined by~(\ref{ec.9}). Finally in Sec. IV some 
particular solutions attainable from the general self-consistent field 
are considered.

\section{An approach for the construction of  self-consistent spherically
 symmetric solutions}
The metric for an arbitrary static, spherically symmetric space-time can be
taken in the form
\begin{eqnarray}
\label{ec.1}
 ds^{2} = e^{2\alpha(r)}dt^{2} - e^{2\beta(r)}dr^{2} - 
        r^{2}(d\theta^{2} + sin^{2} \theta\,\, d\varphi^{2}). 
\end{eqnarray}		
To write Einstein's equations we will use the tetrad formalism
and Cartan structure equations~\cite{Israel}. A convenient orthonormal basis 
for the metric in~(\ref{ec.1}) is
\begin{eqnarray}
\label{ec.2}
 \theta^{(0)} = e^{\alpha} dt , \,\,\, \theta^{(1)} = e^{\beta} dr , \, \,\,
  \theta^{(2)} = r\,\,d\theta , \,\,\, \theta^{(3)} = r\, sin \theta\,
 d\varphi. 
\end{eqnarray}
The stress-energy tensor for a perfect fluid, which fills the space-time,
 for our signature is defined by  
\begin{eqnarray}
\label{ec.3}
  T_{(\mu)(\nu)}\theta^{(\mu)} \otimes \theta^{(\nu)}  =  
  \left[(\mu +\nu)U_{(\mu)} U_{(\nu)} - p g_{(\mu)(\nu)}\right]  
    \mbox{}  \times  \theta^{(\mu)} \otimes \theta^{(\nu)},   
\end{eqnarray}
where $\mu$ and $p$ are the mass-energy density and the pressure of the fluid,
respectively. $U_{(\mu)}$ is its timelike 4-velocity. If we take the four
velocity 
\begin{math}
 {\bf U} = \theta^{(0)}, 
\end{math}
then~(\ref{ec.3}) becomes
\begin{eqnarray}
\label{ec.4}
  T_{(\mu)(\nu)}^{P.F.}=\mu\theta^{(0)}\theta^{(0)}  
     +\,\, p\left( \theta^{(1)}\theta^{(1)}+\theta^{(2)}\theta^{(2)}+
     \theta^{(3)}\theta^{(3)}\right).
\end{eqnarray}
Now,to construct Einstein-Maxwell fields, we must consider Maxwell's
equations and stress-energy tensor of the electromagnetic field. This,
with respect to Eq.~(\ref{ec.2}), is defined by 
\begin{eqnarray}
\label{ec.5}
    T_{(\mu)(\nu)}\theta^{(\mu)}\otimes\theta^{(\nu)} = 
 - \frac{1}{4\pi}[F_{(\mu)(\gamma)}F_{(\nu)}^{ \,\,\,\,\,\, (\gamma)}   
   -\frac{1}{4}g_{(\mu)(\nu)}F_{(\gamma)(\delta)}F^{(\gamma)(\delta)}]
   \theta^{(\mu)}\otimes\theta^{(\nu)}.
\end{eqnarray}
To get its components, of course we must compute $F_{(\mu)(\nu)}$. The
general form for a Maxwell tensor which shares the static and spherical 
symmetries of the space-time is,~\cite{Wald},
\begin{eqnarray}
F_{(\mu)(\nu)} = 2B(r) \, \delta_{[(\mu)}^t \, \delta_{(\nu)]}^r +
              2C(r) \, \delta_{[(\mu)}^\theta \, \delta_{(\nu)]}^\varphi.
\label{ec.6}
\end{eqnarray}
However, our goal in this paper is to obtain superpositions of an 
electromagnetic field and a neutral perfect fluid. To do this, we must 
consider the source-free Maxwell's equations for the self-consistent problem. 
This means that one can set $C$=0 in Eq.~(\ref{ec.6}) since all the cases
with an electric field without sources can be reformulated to the corresponding
cases with a magnetic field (dual to the initial electric one) or mixtures
of both fields (duality rotation). The electromagnetic stress-energy
tensor is the same in all these cases.

In general, in the coordinate basis, the Maxwell tensor can be expressed
in terms of a four-potential covector $A$ in the following way:
\begin{eqnarray}
 dA=F=\frac{1}{2}F_{\mu\nu}dx^{\mu}\wedge dx^{\nu}.
\label{ec.7}
\end{eqnarray}
For the considered Maxwell tensor (with $C$=0) one can take the 4-potential
in the form
\begin{eqnarray}
 A =q\, f(r)\, dt, 
\label{ec.8}
\end{eqnarray}
where $f(r)$ is an arbitrary function of the $r$ coordinate, and $q$ is a 
constant coefficient which is introduced for further use in the 
self-consistent solution (for switching off the electric field).
 
Lastly, we must note that with the help of Eq.~(\ref{ec.8}) the source-free
equations are replaced by an algebraic condition which considerably simplifies
the solution of the self-consistent equations.

\section{Spherically symmetric space-time with a superposition of neutral
 fluid and electric field}
According to the expression~(\ref{ec.8}), we choose the electromagnetic
potential in the form
\begin{equation}
  A = \left\{ \begin{array} {ll}
              \frac{q}{n}\,\, r^n \,dt  &\mbox{if $n\neq0$} \\
              q\,\,ln\,r\,\, dt    &\mbox{if $n=0$}
\end{array}
              \right.  
\label{ec.9}     
\end{equation}

where $n$ is an arbitrary constant. Then
\begin{eqnarray}
 dA=qr^{(n-1)}dr \wedge dt,
\label{ec.10}
\end{eqnarray}
and the Maxwell tensor $F_{\mu\nu}$, in the coordinate basis, has the form
\begin{eqnarray}
 F_{\mu\nu}=2qr^{(n-1)}\delta_{[\mu}^{r} \delta_{\nu]}^{t},
\label{ec.11}
\end{eqnarray}
or in the basis~(\ref{ec.2})
\begin{eqnarray}
 F_{(\mu)(\nu)}=qr^{(n-1)}e^{-\alpha-\beta}\delta_{[\mu}^{1}\delta_{\nu]}^{0}.
\label{ec.12}
\end{eqnarray}
The contravariant density components of~(\ref{ec.11}) are
\begin{eqnarray}
\sqrt{-g}F^{\mu\nu}=qr^{(n-1)}sin\theta \, e^{-\alpha-\beta}
      \delta_{t}^{[\mu}\delta_{r}^{\nu]},
\end{eqnarray}
where $\sqrt{-g}=r^{2}e^{\alpha\beta}sin\theta$. It is clear that the 
source-free Maxwell's equations
\begin{eqnarray*}
 (\sqrt{-g} \, F^{\mu\nu})_{,\nu}=0
\end{eqnarray*}
are satisfied if
\begin{eqnarray}
 r^{n+1}=Ae^{\alpha+\beta},
\label{ec.14}
\end{eqnarray}
where $A$ is a constant of integration. Without any loss of generality, by
re-scaling the $r$ coordinate, we can set $A$=1. In the particular case of an
electric field in the space-time~(\ref{ec.1}), we have locally that
\begin{math}
 F=-E_{r}dt \wedge dr
\end{math}
and we obtain from Eq.~(\ref{ec.10}) that the radial electric field has the
form
\begin{eqnarray}
 E_{r} = q r^{n-1}.
\label{ec.15}
\end{eqnarray}
Now let us write Einstein's equations. By using Cartan exterior forms
calculus one can obtain from the tetrad~(\ref{ec.2}) the next non-trivial
components of the curvature tensor:   
\begin{eqnarray}
      R^{(0)}_{\,\,\,\,\,(1)(1)(0)}&=&
      e^{-2\beta}(\alpha^{'2}+\alpha^{''}-\alpha^{'}\beta^{'}),\label{ec.16} \\
      R^{(0)}_{\,\,\,\,\,(2)(2)(0)}&=&R^{(0)}_{\,\,\,\,\,(3)(3)(0)}=
        \frac{\alpha^{'}}{r}e^{-2\beta},\label{ec.17}  \\
      R^{(1)}_{\,\,\,\,\,(2)(1)(2)}&=&R^{(1)}_{\,\,\,\,\,(3)(1)(3)}=
        \frac{\beta^{'}}{r}e^{-2\beta},\label{ec.18}  \\
 R^{(2)}_{\,\,\,\,\,(3)(2)(3)}&=&\frac{1}{r^{2}}(1-e^{-2\beta}) \label{ec.19}; 
\label{curvatura}
\end{eqnarray}
where the differentiation with respect to $r$ is denoted by  $^{\bf \,\,'}$.  

The stress-energy tensor of the electromagnetic field can be obtained using the
definition~(\ref{ec.5}). The electromagnetic field invariant
\begin{eqnarray*}
 F_{(\mu)(\nu)}F^{(\mu)(\nu)}=-2q^{2}r^{2(n-1)}e^{-2\alpha-2\beta}
\end{eqnarray*}
is negative. This means that we have an electric-type field according to 
Eq.~(\ref{ec.15}). Then
\begin{eqnarray}
\label{ec.20}
      & &T_{(\mu)(\nu)}^{E.F.}\theta^{(\mu)}\otimes\theta^{(\nu)}=
    \frac{q^{2}r^{2(n-1)}e^{-2\alpha-2\beta}}{8\pi} \\  
 & &\times [ \theta^{(0)}\otimes\theta^{(0)}-\theta^{(1)}\otimes\theta^{(1)}+ 
   \theta^{(2)}\otimes\theta^{(2)}+\theta^{(3)}\theta^{(3)}].
     \nonumber
\end{eqnarray}

Thus, from equations~(\ref{ec.4}),~(\ref{ec.16}) --~(\ref{ec.19}) and~(\ref{ec.20}) 
we obtain the non-trivial components of Einstein's equations   
\begin{eqnarray}
 \frac{1}{r^2} - 2 \frac{\beta^{'}}{r} - \frac{e^{2\beta}}{r^{2}} = 
 -\frac{\chi q^{2} r^{2(n-1)} e^{-2\alpha}}{8\pi} - \chi \mu e^{2\beta} , 
\label{ec.21}
\end{eqnarray}
\begin{eqnarray}
 \frac{e^{2\beta}}{r^{2}} -\frac{1}{r^{2}} - 2 \frac{\alpha^{'}}{r}  =
 \frac{\chi q^{2} r^{2(n-1)} e^{-2 \alpha}}{8 \pi} - \chi p e^{2\beta},
\label{ec.22}
\end{eqnarray}
\begin{eqnarray}
\label{ec.23}
\! \alpha^{'}\! \beta^{'}\!\! -\! \alpha{'^{2}}\! -\! \alpha^{''} 
\! -\!\frac{\alpha^{'}}{r}\! +\! \frac{\beta^{'}}{r} = 
 -\frac{\chi q^{2} r^{2(n-1)} e^{-2\alpha}}{8 \pi} 
     -\,\, \chi p e^{2\beta}.    
\end{eqnarray}
 This implies that the self-consistent system consists of equations
~(\ref{ec.14}),~(\ref{ec.21}),~(\ref{ec.22}) and~(\ref{ec.23}).
 
By subtracting equations~(\ref{ec.22}) and~(\ref{ec.23}) and taking into
account Eq.~(\ref{ec.14}) (with $A$=1), we obtain
\begin{eqnarray}
     (e^{2\alpha})^{''}-(n+1)\frac{(e^{2\alpha})^{'}}{r}-
      2(n+2)\frac{e^{2\alpha}}{r^{2}}=  
      \frac{\chi q^{2}}{2\pi}r^{2(n-1)}-2r^{2n},
\end{eqnarray}
which implies
\begin{eqnarray}
 e^{2\alpha} = \frac{Q^{2} r^{2n}}{n^{2}-3n-2} + 
    \frac{r^{2(n+1)}}{2-n^{2}}+Ar^{I_{1}(n)} + Br^{I_{2}(n)} ,
\label{ec.25}
\end{eqnarray}
where 
\begin{eqnarray}
I_{1}(n) = \frac{1}{2}\left(n+2+\sqrt{(n+2)(n+10)}\right) ,
\label{ec.26}
\end{eqnarray}
\begin{eqnarray}
  I_{2}(n) = \frac{1}{2}\left(n+2-\sqrt{(n+2)(n+10)}\right).        
\label{ec.27}
\end{eqnarray}

$A$ and $B$ are constants of integration and $Q^{2}=\chi q^{2}/4\pi$. From 
equations~(\ref{ec.26}),~(\ref{ec.27}) it follows that the possible range 
for values of $n$ is $n\leq-10$ or $n\geq-2$ (for nonvanishing constants $A$ 
and $B$). On the other hand, 
equation~(\ref{ec.14}) yields
\begin{eqnarray}
  e^{2\beta}=\frac{r^{2(n+1)}}{\frac{Q^{2} r^{2n}}{n^{2}-3n-2} + 
    \frac{r^{2(n+1)}}{2-n^{2}}+Ar^{I_{1}(n)} + Br^{I_{2}(n)}}.
\end{eqnarray}
Now substituting the functions $\alpha$ and $\beta$ in equations~(\ref{ec.21}) 
and~(\ref{ec.22}), and performing a few algebraic 
manipulations, we find that for the energy density of the fluid,
\begin{eqnarray}
\lefteqn{\chi \mu(r) = \frac{Q^{2} (n+1)(4-n)}{2(n^{2} -3n -2)r^{4}} -
          \frac{n^{2}-1}{(2-n^{2})r^{2}} + r^{-2(n+2)}}   \\ 
    \mbox{} \nonumber \\  
   & &         \times   \left\{A(2n + 1 - I_{1})r^{I_{1}(n)} + 
             B(2n + 1 - I_{2})r^{I_{2}(n)} \right\} ,  \nonumber
\end{eqnarray}
and for its pressure,
\begin{eqnarray}
 \chi p(r) = \frac{Q^{2} n(n+1)}{2(n^{2}-3n-2)r^{4}} + 
        \frac{(n+1)^{2}}{(2-n^{2})r^{2}} +  \\  
         \mbox{} \nonumber \\
        r^{-2(n+2)} \times  \left\{A(I_{1}+1)r^{I_{1}(n)} +
               B(I_{2}+1)r^{I_{2}(n)} \right\}.  \nonumber
\end{eqnarray}
Thus we arrive at the space-time metric
\begin{eqnarray}
 ds^{2}=e^{2\alpha}dt^{2}\!\!-\!\!r^{2(n+1)}e^{-2\alpha}dr^{2}\!\!-\!\! 
    r^{2}(d\theta^{2}\!+\!sin^{2}\theta d\varphi^{2}), 
\label{ec.29}
\end{eqnarray}
where $e^{2\alpha}$ is given by Eq.~(\ref{ec.25}).

\section{Limiting cases and the Weyl tensor for the found space-time}
First we will consider the case when the electric field is switched off. 
This means that $q=0$ in our solution. Thus we obtain an analytical
solution which depends on an arbitrary parameter $n$ for the case where the
only source of the gravitational field is a perfect fluid, and the geometry
is static and possesses spherical symmetry. When $n=-1$ we have the Kottler
solution~\cite{Kottler}, and when $A=0$ the Schwarzschild geometry~\cite{Sch}.
 Now, if also $A=B=0$, the solution becomes
\begin{eqnarray}
 ds^{2}=r^{2(n+1)}dt^{2}\! -\!(2\!-\!n^{2})dr^{2}\!-\!
          r^{2}\left(d\theta^{2}\!+\!sin^{2}\theta d\varphi^{2}\right) ,
\end{eqnarray}
with
\begin{eqnarray}
          \chi\mu = \frac{1-n^{2}}{(2-n^{2})r^{2}}   
\end{eqnarray}
and
\begin{eqnarray}
          \chi p = \frac{(n+1)^{2}}{(2-n^{2})r^{2}} .
\end{eqnarray}
If the fluid obeys a $\gamma$-law equation of state, i.e., its $\mu$ and $p$ 
are related by an equation of the form 
\begin{eqnarray}
 p=(\gamma-1)\mu,
\end{eqnarray}
where $\gamma$ is a constant (which, for physical reasons~\cite{Collins} 
satisfies the inequality $1\leq\gamma\leq2$), the constant $\gamma$ may be 
expressed as
\begin{eqnarray}
 \gamma = \frac{2}{1-n}.
\end{eqnarray}
Therefore, the possible range of values of $n$ (when $A=B=Q=0$)  is 
$-1\leq n\leq0$. In the present case, when $n=-1/2$ we get exactly the Klein 
metric for a static spherically symmetric distribution of incoherent 
radiation~\cite{Cataldo2,Klein}, and when $n=0$ a space-time filled with stiff 
matter.

Returning to the case where the electric field is present, we let $n=-1$, so
that our solution represents the well known {\bf exterior electrovacuum} of
charged Kottler space-time~\cite{Cataldo2}. Moreover, if $A=0$ we obtain the
Reissner-Nordstr\"{o}m solution~\cite{Reissner}. 

When $n=-2$ the static equation can be written as $\mu+3p=0$, and when 
$n=-\frac{1}{2}$ we obtain a generalization of the Klein metric where a
superposition of an  electric field and a perfect fluid is present. Another 
generalization of the Klein metric  has been obtained in~\cite{Cataldo2}, 
where the static perfect fluid is charged.

On the other hand, the only curvature singularity of the metric~(\ref{ec.29})
is located at $r=0$ and the horizons at $e^{2\alpha}=0$, forming purely 
coordinate singularities.

Now, let us consider a special case of the found metric setting $A=B=0$. Then
\begin{eqnarray}
\label{ec.35}
\lefteqn{ds^{2}=\left(\frac{Q^{2}r^{2n}}{n^{2}-3n-2}+
       \frac{r^{2(n+1)}}{2-n^{2}}\right)dt^{2}-}  \\
       \mbox{} \nonumber \\
   & &    r^{2(n+1)}\left(\frac{Q^{2}r^{2n}}{n^{2}-3n-2}+
      \frac{r^{2(n+1)}}{2-n^{2}}\right)^{-1}dr^{2}  
       -r^{2}(d\theta^{2}+sin^{2}\theta d\varphi^{2}), \nonumber
\end{eqnarray}
\begin{eqnarray}
 \chi \mu=\frac{Q^{2}(n+1)(4-n)}{2(n^{2}-3n-2)r^{4}}+
         \frac{1-n^{2}}{(2-n^{2})r^{2}},
\end{eqnarray}
and
\begin{eqnarray}
 \chi p=\frac{Q^{2}n(n+1)}{(n^{2}-3n-2)r^{4}}+
        \frac{(n+1)^{2}}{(2-n^{2})r^{2}}. 
\end{eqnarray}
In this case we can characterize the behaviour of the fluid as follows. At
\begin{eqnarray*}
 r_{1}=\sqrt{\frac{Q^{2}(n^{2}-2)(n-1)}{(n^{2}-3n-2)(2n+1)}}
\end{eqnarray*}
as an incoherent radiation ($\mu=3p$), at
\begin{eqnarray*}
 r_{2}=\sqrt{\frac{Q^{2}(2-n^{2})(2-n)}{2n(n^{2}-3n-2)}}
\end{eqnarray*}
as a stiff matter ($\mu=p$),at 
\begin{eqnarray*}
 r_{3}=\sqrt{\frac{Q^{2}n(n^{2}-2)}{2(n^{2}-3n-2)(n+1)}}
\end{eqnarray*}
as a dust, and at
\begin{eqnarray*}
 r_{4}=\sqrt{\frac{Q^{2}(n^{2}-2)}{n^{2}-3n-2}}
\end{eqnarray*}
as an electrified de Sitter space-time ($\mu=-p$). The latter spherical surface
is the horizon for the resulting static spacetime~(\ref{ec.35}). For their
existence, it must be true that all of the values of the spherical surfaces
have to be positive for a determined value of $n$. Thus, for example, the 
horizon exists only if $\frac{3-\sqrt{17}}{2}<n<\sqrt{2}$ or 
$n>\frac{3+\sqrt{17}}{2}$ or $n<-\sqrt{2}$. 
 
Lastly, we find that the Weyl coefficients {\large  ${\psi_{a}}$} for the 
general solution~(\ref{ec.29}), with respect to the complex null
tetrad
\begin{eqnarray*}
   \Theta^{(0)}&=&\sqrt{\frac{1}{2}}(\theta^{(0)}-\theta^{(1)}), \\
   \Theta^{(1)}&=&\sqrt{\frac{1}{2}}(\theta^{(0)}+\theta^{(1)}), \\ 
   \Theta^{(2)}&=&\overline{\Theta^{3}}=
       \sqrt{\frac{1}{2}}(\theta^{(0)}+\theta^{(1)}),
\end{eqnarray*}
are
\begin{eqnarray*}
 \psi_{0}&=&\psi_{1}\,\,\,\,=\,\,\,\,\psi_{3}\,\,\,\,=\,\,\,\,\psi_{4}\,\,\,\,=0 \\
 \mbox{} \\
 \psi_{2}&=&-\frac{Q^{2}}{6r^{4}}+
   \frac{(n^{2}-1)}{3(n^{2}-2)r^{2}}- \\
 & & \frac{A}{12}I_{1}(n)\left\{ I_{1}(n)+3n+4 \right\}r^{(I_{1}(n)-2n-4)}- \\
 & & \frac{B}{12}I_{2}(n)\left\{ I_{2}(n)+3n+4 \right\}r^{(I_{2}(n)-2n-4)}.
\end{eqnarray*}
We see that in general, as is well known for any static spherically 
symmetric space-time, the solution~(\ref{ec.29}) belongs to
Petrov type D, degenerating locally to the conformally flat case (type 0)
when $\psi_{2}=0$. For the special case~(\ref{ec.35}), type 0 takes place
for the spherical surface
\begin{eqnarray*}
 r=\sqrt{\frac{Q^{2}(n^{2}-2)}{2(n^{2}-1)}}
\end{eqnarray*}
and asymtotically when r $\rightarrow \infty$.

\section*{Conclusions}
 From some properties of static, spherically symmetric space-times, it has 
been shown that one can choose the electromagnetic four-potential in the 
form~(\ref{ec.9}). This choice allows one to construct the two-parameter set 
of metrics~(\ref{ec.29}), which are solutions of the Einstein-Maxwell equations.
The results obtained in Sec. III lead us to conclude that the obtained 
self-consistent space-time describes a field, where is 
a point charge located at the origin ($r=0$). It is important to remark that 
the general solution~(\ref{ec.29}) is valid only outside of the electric 
source since are
considered the sourceless Maxwell equations. In fact we have a neutral fluid
and an electric field given by~(\ref{ec.15}). This leads us to think that the 
electric field can be interpreted as a superposition of a spherically
symmetric coulomb field (external electromagnetic field) and another field 
caused by ``polarization'' of the fluid, given by 
$E_{f}=\frac{q}{r^{2}}(r^{n+1}-1)$ which vanishes when $n=-1$. This is in 
accordance with the found solution since, if $n=-1$, the fluid is ``switched 
off'' and we obtain the Kottler-Reissner-Nordstr\"{o}m electrovacuum solution.
Also, if $A=0$, we obtain the charged black hole (Reissner-Nordstr\"{o}m solution).

Finally we can say that the self-consistent solution depends on the function
$f(r)$ which appears in~(\ref{ec.8}) for the electromagnetic potential. Also 
setting $q=0$ one obtains an exact solution for Einstein's equations.

\section*{Acknowledgments}
We would like to thank the referee for his helpful comments. We also want to 
thank P. Minning and H. Miranda for their help in the redaction of the 
manuscript.

This work was supported in part by Direcci\'on de Promoci\'on y Desarrollo de 
la Universidad del B\'\i o-B\'\i o through Grants \# 942305-1C, 951105-1 
and in part by Direcci\'on de Investigaci\'on de la Universidad de Concepci\'on
through Grant P.I. \# 94.11.09-1.
\end{document}